\documentclass[%
 reprint,
 amsmath,amssymb,
 aps,
]{revtex4-2}

\usepackage{graphicx}
\usepackage{dcolumn}
\usepackage{bm}
\usepackage{xcolor}
\usepackage{dsfont}
\usepackage{pgfplots}
\pgfplotsset{compat=1.15}
\usepackage{mathrsfs}
\usetikzlibrary{arrows}
\newcommand{\ket}[1]{| {#1} \rangle} 

\usepackage[utf8]{inputenc}
\usepackage[T1]{fontenc}
\usepackage[english,serbian]{babel}
\usepackage[OT1]{fontenc}
\usepackage{mathrsfs}
\usepackage{graphicx}
\usepackage{amssymb}
\usepackage{amsbsy}
\usepackage{latexsym}
\usepackage{mathtools}
\usepackage{titlesec}
\usepackage{amsmath}
\usepackage{color}
\usepackage[utf8]{inputenc}
\usepackage{lipsum}
\usepackage{fontenc}
\usepackage{bbold}
\usepackage{tikz-cd}

\usepackage{chngcntr}
\numberwithin{equation}{section}
\renewcommand{\theequation}{\arabic{section}.\arabic{equation}}

\def\dj{d\kern-0.4em\char"16\kern-0.1em}

\def\diff{\textrm{d}}
\def\Diff{\textrm{D}}

\def\tr{{\rm Tr}}

\def\dj{d\kern-0.4em\char"16\kern-0.1em}
\def \Dj {\mbox{\raise0.3ex\hbox{-}\kern-0.4em D}}

\begin{document}
\definecolor{ududff}{rgb}{0.30196078431372547,0.30196078431372547,1.}
\definecolor{bcduew}{rgb}{0.7372549019607844,0.8313725490196079,0.9019607843137255}
\preprint{APS/123-QED}

\title{Holographic Aspects of Even-Dimensional Topological Gravity}

\author{Du\v{s}an \Dj or\dj evi\'{c} and Dragoljub Go\v{c}anin}
\affiliation{Faculty of Physics, University of Belgrade, Studentski Trg 12-16, 11000 Belgrade, Serbia}

\email{djdusan@ipb.ac.rs}

\selectlanguage{english}


\begin{abstract}

In an odd-dimensional spacetime, gravity can be formulated as a proper gauge theory based on the Chern-Simons action for a suitable gauge group. Performing dimensional reduction, one obtains, as an effective theory, Chamseddine's even-dimensional topological gravity with the reduced gauge symmetry. This theory involves a multiplet of scalar fields that appear as a result of the dimensional reduction, and it is topological in the sense that its action does not depend on the metric. Focusing primarily on the four-dimensional case, we use the holographic dictionary to compute one-point correlation functions of the relevant boundary operators and find that the spin-current can have a nonzero expectation value in the dual quantum field theory. We also consider the generalized holographic Weyl anomaly and find that it vanishes. Finally, we propose a way of computing two-point correlation functions using the gravitational Wilson lines.

\end{abstract}

\maketitle

\selectlanguage{english}

\section{Introduction}

Holographic duality is one of the tenets of modern quantum gravity research, the most prominent example being the AdS/CFT correspondence \cite{Maldacena:1997re} - a conjectured duality between a theory of quantum gravity in $\mathrm{(D+1)}$-dimensional asymptotically anti-de Sitter (AdS) spacetime (the bulk) and a conformal field theory (CFT) that resides on the $\mathrm{D}$-dimensional asymptotic boundary. The standard version of this duality works in a regime where the gravity is well approximated by a semi-classical theory while the dual CFT is strongly coupled. In particular, the holographic dictionary \cite{Witten:1998qj} states that bulk fields are associated with dual CFT operators, and this matching is usually done by using the most ``divergent'' component of the bulk field as the source for the CFT operator. Moreover, the gravitational partition function in the saddle point approximation corresponds to the partition function of the dual CFT (if the appropriate boundary conditions are imposed). In the context of string theory,  initial considerations established an equivalence between superstring (supergravity) theory on AdS$_\mathrm{5}\times \mathrm{S^5}$ and $\mathcal{N}=4$ supersymmetric Yang-Mills theory in $\mathrm{D}=4$. Soon after this pioneering work, the holographic duality 
was successfully applied to various models of gravity and opened a new way of studying condensed matter systems. One notable example is the holographic relation between Jackiw–Teitelboim (JT) gravity \cite{Engelsoy:2016xyb} and the Sachdev–Ye–Kitaev (SYK) model \cite{Sachdev:2010um, Maldacena:2016hyu}.
The correspondence was also generalised to the case of non-Riemannian geometries, where nonzero torsion plays an important role in sourcing the boundary spin-current \cite{Banados:2006fe}. Based on these previous considerations, in this paper, we focus on even-dimensional gauge theories of gravity in the bulk and their holographic description in terms of boundary correlation functions. 

Unlike the Standard Model of particle physics, General Relativity (GR) is not formulated as a proper gauge theory. There are, however, gravity theories that are of this type; those were formulated by Chamseddine \cite{Cham1, Cham2} in any number of spacetime dimensions. In an odd-dimensional spacetime, they coincide (up to a boundary term) with the Chern-Simons (CS) action for a suitable gauge group. The holographic description of 5-dimensional CS gravity was studied in \cite{Banados:2005rz, Banados:2006fe, Cvetkovic:2017fxa}, and the role of torsion in the context of AdS/CFT was addressed in \cite{Gallegos:2020otk}. It was argued that torsion could be used to introduce spin-current degrees of freedom at the boundary useful for describing the hydrodynamics of spin systems - in the case of \cite{Gallegos:2020otk}, the resulting theory was defined on a 4-dimensional boundary.
On the other hand, in an even-dimensional spacetime, besides a gauge connection, a multiplet of scalar fields has to be introduced. However, all those theories have a common property that their action can be written only using differential forms and wedge products without the Hodge dual, and in that sense, we regard them as topological (though they can have local propagating degrees of freedom \cite{Mignemi:1998us}). For future reference, we will dub the even-dimensional ones - Chamseddin's topological gravity (CTG) theories.

Furthermore, it was demonstrated in \cite{Cham2} that 5-dimensional CS gravity action with conformal $SO(4,2)$ gauge symmetry can be dimensionally reduced by Kaluza-Klein compactification to 4-dimensional CTG action with $SO(3,2)$ gauge symmetry. In this paper, we will use this fact to derive the holographic dual of 4-dimensional CTG theory with a nontrivial spin-tensor defined on a 3-dimensional boundary (and, in addition, on any odd-dimensional manifold with a dimension greater than three). We stress the role of the bulk scalars and show that non-vanishing torsion in the bulk is not a necessary condition for having a spin-current on the boundary. One could also take a different point of view and look at the 4-dimensional CTG action as a straightforward generalisation of the 2-dimensional BF theory that produces the equations of motion of JT gravity (section \ref{JT}). The 2-dimensional BF gravity can also be obtained from the 3-dimensional CS gravity via dimensional reduction, a fact that has been used previously in the literature on holography \cite{Ebert:2022ehb}. As explained in \cite{Ebert:2022ehb}, in order to get an interesting boundary theory, one has to deform the original action by introducing appropriate boundary terms. As a more technical part of this paper, we perform this kind of deformation of the CTG theory so that the holographically motivated boundary conditions are satisfied. 

The plan of this paper is the following. In section \ref{redukcija}, we use the relation between 5-dimensional CS gravity and 4-dimensional CTG to identify the Fefferman-Graham expansion of the bulk fields. In section \ref{struje}, we calculate the holographic currents and analyze the obtained results. Moreover, we analyze some solutions of the bulk equations of motion and discuss some of their properties in the context of holography. In section \ref{JT}, as a way of confirmation, we apply our procedure to the case of 2-dimensional BF theory, whose equations of motion yield JT gravity. Section \ref{wlline} is devoted to the study of line defects in the bulk, interpreted as heavy particles. Finally, section \ref{kraj} contains our conclusions and outlook. The summary of notation, conventions and the algebraic setup is presented in Appendix A. Derivation of the Fefferman-Graham gauge is given in Appendix B, and the generalization of the results obtained for the 4-dimensional CTG to any even number of spacetime dimensions can be found in Appendix C.

\section{Dimensional reduction and the holographic ansatz} \label{redukcija}

Here we give a short account of the 5-dimensional CS gravity and its dimensional reduction, see \cite{Cham2, Dordevic:2022ruk} for more details. Throughout, we use the first-order formalism where the vielbein, $\hat{E}$, and the spin-connection, $\hat{\Omega}$, are treated as independent fields. The notation, conventions and some background algebra can be found in Appendix A. Suppressing the wedge product, the action for the 5-dimensional CS gravity is given by 
\begin{align}\label{CS5D}
&S^{(5\text{D})}_{\text{CS}}= \frac{k}{8}\int_{\mathcal{M}_{5}}\varepsilon_{\mathbb{A}\mathbb{B}\mathbb{C}\mathbb{D}\mathbb{E}}\bigg(\frac{1}{l}\hat{R}^{\mathbb{A}\mathbb{B}}\hat{R}^{\mathbb{C}\mathbb{D}}\hat{E}^{\mathbb{E}}\nonumber\\
&+\frac{2}{3l^3}\hat{R}^{\mathbb{A}\mathbb{B}}\hat{E}^{\mathbb{C}}\hat{E}^{\mathbb{D}}\hat{E}^{\mathbb{E}}
+\frac{1}{5l^5}\hat{E}^{\mathbb{A}}\hat{E}^{\mathbb{B}}\hat{E}^{\mathbb{C}}\hat{E}^{\mathbb{D}}\hat{E}^{\mathbb{E}}\bigg),
\end{align} 
where $\hat{R}=\mathrm{d}\hat{\Omega}+\hat{\Omega}^{2}$ is the curvature $2$-form, $k$ is a dimensionless constant (the CS level), and $l$ is the appropriate length scale; henceforth, we set $l=1$. Up to a boundary term, the action (\ref{CS5D}) is invariant under the conformal gauge group $SO(4,2)$. 
The equations of motion for the independent fields $\hat{E}^{\mathbb{A}}$ and $\hat{\Omega}^{\mathbb{A}\mathbb{B}}$ are
\begin{align}
\varepsilon_{\mathbb{A}\mathbb{B}\mathbb{C}\mathbb{D}\mathbb{E}}\big(\hat{R}^{\mathbb{A}\mathbb{B}}+\hat{E}^\mathbb{A}\hat{E}^\mathbb{B}\big)\big (\hat{R}^{\mathbb{C}\mathbb{D}} +\hat{E}^\mathbb{C} \hat{E}^\mathbb{D}  \big)&=0,\\
\varepsilon_{\mathbb{A}\mathbb{B}\mathbb{C}\mathbb{D}\mathbb{E}}\hat{T}^\mathbb{A}\big( \hat{R}^{\mathbb{B}\mathbb{C}}+\hat{E}^\mathbb{B}\hat{E}^\mathbb{C}\big)&=0.    
\end{align}
Note that the torsion $2$-form $\hat{T}=\mathrm{d}\hat{E}+\hat{\Omega}\hat{E}$ does not necessarily vanish on-shell.

As demonstrated in \cite{Cham2}, by compactifying one spatial dimension (the one corresponding to the spacetime index 4) into a circle, the 5-dimensional CS gravity action (\ref{CS5D}) reduces, up to a boundary term (which will be important later on), to the following 4-dimensional CTG action with $SO(3,2)$ gauge symmetry, 
\begin{align}\label{ovde}
S_{CTG}&=\kappa \int_{\mathcal{M}_{4}} \varepsilon_{\hat{A}\hat{B}\hat{C}\hat{D}\hat{E}}\hat{\Phi}^{\hat{A}} \hat{F}^{\hat{B}\hat{C}} \hat{F}^{\hat{D}\hat{E}},
\end{align}
where $\hat{F}$ stands for the $SO(3,2)$ field strength 2-form and $\hat{\Phi}$ is a multiplet of spacetime scalars that
appear after dimensional reduction. The parameter $\kappa$ is defined in terms of the CS level $k$ and the compactification radius; the radius has to be small enough so that we can ignore the higher Kaluza-Klein modes. 
The effective 4-dimensional CTG action (\ref{ovde}) describes our bulk theory of gravity and it will be our starting point for the holographic analysis; all bulk fields are denoted by a hat symbol. 

The equations of motion obtained by varying action (\ref{ovde}) with respect to $\hat{\Phi}^{\hat{A}}$ and the full $SO(3,2)$ connection $\hat{\Omega}^{\hat{A}\hat{B}}$ are given by
\begin{align}\label{istakaocs}
\varepsilon_{\hat{A}\hat{B}\hat{C}\hat{D}\hat{E}}\hat{F}^{\hat{A}\hat{B}}\hat{F}^{\hat{C}\hat{D}}=0,\\
\varepsilon_{\hat{A}\hat{B}\hat{C}\hat{D}\hat{E}}\hat{F}^{\hat{A}\hat{B}}\Diff \hat{\Phi}^{\hat{C}}=0.
\end{align}
where $\Diff$ stands for the $SO(3,2)$ covariant derivative. The $SO(3,2)$ gauge group index is decomposed as $\hat{A}=(A,5)$ where again $A=0,1,2,3$ is the standard Lorentz index. Using the field strength components, $\hat{F}^{AB}=\hat{R}^{AB}+\hat{e}^A\hat{e}^B$ and $\hat{F}^{A5}=\hat{T}^{A}$, 
the previous two equations can be cast \cite{Mignemi:1998us} in a more explicit form, 
\begin{align}\label{jna1}
\varepsilon_{ABCD}\big(\hat{R}^{AB}+\hat{e}^A\hat{e}^B   \big)\big (\hat{R}^{CD} +\hat{e}^C \hat{e}^D  \big)&=0,\\\label{jna2}
\varepsilon_{ABCD}\hat{T}^A\big( \hat{R}^{BC}+\hat{e}^B\hat{e}^C  \big)&=0,\\\label{jna3}
\varepsilon_{ABCD}\big( \hat{R}^{BC}+\hat{e}^B\hat{e}^C \big)\big( \Diff \hat{\phi}^A-\hat{\varphi}\hat{e}^A \big)&=0, \\\nonumber
\varepsilon_{ABCD}\big( 2\hat{T}^B(\Diff \hat{\phi}^A-\hat{\varphi}\hat{e}^A)&\\ \label{jna4}
+(\hat{R}^{AB}+\hat{e}^A\hat{e}^B  )(\diff\hat{\varphi}-\hat{\phi}^E \hat{e}_E)\big )&=0,
\end{align}
where $\hat{R}^{AB}=\mathrm{d}\hat{\omega}^{AB}+\hat{\omega}^{A}_{\;\;C}\hat{\omega}^{CB}$is the bulk curvature and  $\hat{T}^{A}=\mathrm{d}\hat{e}^{A}+\hat{\omega}^{AB}\hat{e}_{B}$ is the bulk torsion. 

Since boundary terms may play an essential role in holography, we will start from the action (\ref{ovde}) instead of making a direct dimensional reduction of all the results pertaining to the original 5-dimensional CS gravity case. Nevertheless, those results that are independent of the boundary terms can be obtained directly by using the reduction prescription. In particular, this is true for the Fefferman-Graham (FG) expansion of the bulk fields - an expansion of the bulk fields organized in powers of the radial coordinate $\rho$. For CS gravity, the FG expansion is found in \cite{Banados:2006fe}. The fact that on-shell action for this theory is IR divergent means that appropriate regularisation and renormalisation procedure has to be imposed. The FG expansion is finite, as opposed to more generic situations. Actually, the CS gravity has to be considered separately from other generic Lovelock gravity theories, as the equations of motion are degenerate, and the theory is (in a sense that has already been explained) topological \cite{Aksteiner:2015uxw}.   
In general, the structure of the FG expansion is based on the diffeomorphism invariance, gauge invariance and the equations of motion of the bulk theory. Since the dimensional reduction of the 5-dimensional CS gravity action to 4-dimensional CTG action consistently extends to the equations of motion and the symmetry structure of the two theories, we can directly reduce the FG expansions from the 5-dimensional CS gravity \cite{Banados:2006fe} and write down the asymptotic FG expansion of the fields appearing in our 4-dimensional CTG theory. 

The asymptotic boundary is located at $\rho=0$. Boundary fields (written without the hat symbol) are finite and do not have $\mathrm{d}\rho$ component.
The index 1 corresponds to the radial coordinate $\rho$, and the Lorentz index is decomposed as $A=(a,1)$, with $a=0,2,3$ being the boundary index. The asymptotic expansions (they are finite as for the 5-dimensional CS gravity from which they are derived) of the bulk fields are given by 
\begin{align}\label{e}
&\hat{e}^1=-\frac{\diff \rho}{2\rho},\hspace{0.4cm} 
\hat{e}^a=\frac{1}{\sqrt{\rho}}(e^a+\rho k^a),\\
\label{omega}
&\hat{\omega}^{a1}=\frac{1}{\sqrt{\rho}}(e^a-\rho k^a),\hspace{0.4cm}
\hat{\omega}^{ab}=\omega^{ab}, \\
\label{fi}
&\hat{\phi}^1=\frac{1}{\sqrt{\rho}}(\varphi-\rho\psi), \hspace{0.4cm}\hat{\phi}^a=\phi^a, \\
\label{varfi}
&\hat{\varphi}=\frac{1}{\sqrt{\rho}}(\varphi+\rho\psi).
\end{align} 
This holographic ansatz has to satisfy the 4-dimensional bulk equations of motion (\ref{jna1})-(\ref{jna4}), which gives us a set of constraints on the boundary fields (see subsection \ref{geom}).

\section{Holographic currents}\label{struje}

The typical situation with gravity in asymptotically AdS spacetimes is the following. Integration of the Lagrangian density down to $\rho=0$ introduces divergences in the on-shell action. In order to be able to interpret the on-shell bulk gravity action as the generating function for the dual CFT, one has to perform holographic renormalization. This is not surprising, as renormalization plays an important role in quantum field theory (QFT), the only difference being that QFT has to be renormalized in the UV regime, while the bulk gravity is IR divergent. An important aspect of the AdS/CFT correspondence is that it relates the IR scales of gravity in the bulk and the UV scales of the corresponding CFT at the boundary. The renormalization is achieved by putting an IR cutoff on the gravitational side and adding appropriate boundary counterterms that do not influence classical equations of motion but lead to a finite expression for the boundary correlation functions. 
Another important aspect of the boundary terms is their connection to the variation principle. In order to talk about dynamics, one has to define a set of boundary conditions and, in some cases, deform the theory by adding appropriate boundary terms. Note that action (\ref{ovde}) vanishes on-shell, and therefore we have to be careful when setting up the variational principle in order to get an interesting boundary QFT. This is analogous to the situation with the BF formulation of JT gravity \cite{Ebert:2022ehb, Iliesiu:2019xuh} (see also section \ref{JT}). 

We follow the procedure developed in \cite{Banados:2006fe}. The variation of the 4-dimensional CTG action (\ref{ovde}) is given by 
\begin{equation}
\kappa \int_{\mathcal{M}_{4}}\varepsilon_{\hat{A}\hat{B}\hat{C}\hat{D}\hat{E}}\big(\delta \hat{\Phi}^{\hat{A}}\hat{F}^{\hat{B}\hat{C}}\hat{F}^{\hat{D}\hat{E}} +2\hat{\Phi}^{\hat{A}}\delta \hat{F}^{\hat{B}\hat{C}}\hat{F}^{\hat{D}\hat{E}}\big). 
\end{equation}
Decomposing indices, we get
\begin{align}
&\kappa\int_{\mathcal{M}_{4}}\varepsilon_{ABCD}\Big(\delta\hat{\varphi}\hat{F}^{BC}\hat{F}^{DE}+2\delta\hat{\phi}^{A}\hat{T}^{B}\hat{F}^{CD}\\
&+2\hat{\varphi}\delta \hat{F}^{AB}\hat{F}^{CD}+4\hat{\phi}^A\delta \hat{T}^B\hat{F}^{CD}
+4\hat{\phi}^A\hat{T}^B\delta\hat{F}^{CD}\Big).\nonumber
\end{align}
After some partial integration, putting the variation on-shell yields 
\begin{align}\nonumber
\delta S\; |_{\mathrm{on-shell}} =\kappa &\int_{\partial\mathcal{M}_{4}}\varepsilon_{ABCD}\Big(2\hat{\varphi}\delta\hat{\omega}^{AB}\hat{F}^{CD}\\&+4\hat{\phi}^{A}\delta\hat{e}^B \hat{F}^{CD}+4\hat{\phi}^{A}\hat{T}^B\delta\hat{\omega}^{CD}\Big).
\end{align}
There is no variation of the fields $\hat{\varphi}$ and $\hat{\phi}^{A}$ field since there are no derivatives of these fields in the action. We now use the asymptotic expansions (\ref{e})-(\ref{varfi}) to organize the action in powers of $\rho$. In general, we should care only about those terms that are of order $\rho^{0}$. This is because the renormalization theorem \cite{Cvetkovic:2017fxa} claims that we can rewrite terms that contain nonzero powers of $\rho$ as $\delta(\dots)$ and thus those terms can always be compensated by adding counterterms to the original action. However, in our case, we can even check that terms that contain nonzero powers of $\rho$ actually vanish, which leaves us only with finite terms. They are given by 
\begin{align}
    \delta S|_{\mathrm{on-shell}}=&4\kappa \int_{\partial\mathcal{M}_{4}}\varepsilon_{abc}\times\\
    \times\Big(&\delta k^a\big(-2\varphi(R^{bc}+4e^{b}k^c)-4\phi^b T^c\big)\nonumber\\
    +&\delta e^a\big ( 2\psi(R^{bc}+4e^bk^c)+4\phi^b\Diff k^c\big) \nonumber\\
   +&\delta \omega^{ab}\big ( -2\varphi \Diff k^c+2\psi T^c-2\phi^ce^dk_d\big ) \Big),\nonumber
\end{align}
where now $\mathrm{D}$ stands for the Lorentz covariant derivative.

Boundary fields $e^a$ and $\omega^{ab}$ couple to the field theory stress-energy tensor $\tau_{a}$ and the spin-tensor $\sigma_{ab}$, respectively, while fields $\varphi$ and $\phi^a$, if their variation would be present in the above expression, would couple to certain operators, $o_{\varphi}$ and $o_{a}$, in the boundary QFT. Yet, the variations of $\varphi$ and $\phi^a$ do not appear in the last expression. Moreover, there is no obvious choice for what $k^a$ should couple with. This motivates us to define the boundary conditions such that only fields $e^a$, $\omega^{ab}$, $\varphi$ and $\phi^{a}$ are fixed at the boundary, as they are interpreted as boundary sources in AdS/CFT. This is different from the standard choice of boundary conditions where one fixes the full gauge connection at the boundary. Note also that in the case of asymptotically AdS spacetimes, due to divergences present in the asymptotic expansions at $\rho=0$, it is hard to give a physical meaning to the standard boundary conditions. 

This $k^a$ is not determined by the bulk equations of motion, and we can therefore add appropriate boundary terms to move the variation from $k^a$ to other fields, thus deforming the original theory. New, deformed theory, has nonzero on-shell action. We illustrate this for one of the terms, namely 
\begin{align}\nonumber
\int_{\partial\mathcal{M}_{4}}\varepsilon_{abc}\delta k^a \varphi R^{bc}=\int_{\partial\mathcal{M}_{4}} \varepsilon_{abc}\big(\delta(k^a\varphi R^{bc})-&\delta\varphi k^a R^{bc}\\-\Diff(k^a\varphi)\delta \omega^{bc}\big)&.
\end{align}
The boundary term $\diff(\varepsilon_{abc}\varphi k^a \omega^{bc})$ is discarded as $\partial^{2}\mathcal{M}_{4}=\emptyset$. In total, the boundary term that we have to add in order to respect the holographic boundary conditions is 
\begin{equation}\label{GHY1}
S_{GHY}=8\kappa \int_{\partial\mathcal{M}_4} \varepsilon_{abc}\left ( \varphi k^aR^{bc} +2\varphi k^ak^be^c+2k^a\phi^bT^c   \right ).
\end{equation}
This term is finite and can be thought of as a generalised on-shell Gibbons-Hawking-York (GHY) term, as explained in \cite{Erdmenger:2022nhz}.
For convenience, we should relate the constant $\kappa$ to Newton's constant $G$. Since the CTG action is not the standard Einstein-Hilbert action, it is not possible to directly see the relation.
However, motivated by the fact that CTG action contains the Einstein-Hilbert term, multiplied by the field $\hat{\varphi}$, 
we will introduce the constant $G$ using the following relation 
\begin{equation}\nonumber   
4\kappa=\frac{1}{16\pi G}.
\end{equation}
Note that this fact is also used in the MacDowell-Mansouri-Chamseddine-Stelle-West (MMCSW) approach to 4-dimensional gravity with negative cosmological constant \cite{MacDowell-Mansouri,stelle-west}, although this formalism is incompatible with our analysis that relies heavily on the local $SO(3,2)$ symmetry, which is broken in the MMCSW.  

The final expression for the variation of the modified CTG action is, therefore, given by 
\begin{align}\nonumber
\delta S_{\mathrm{mod}}=&\frac{1}{16\pi G}\int_{\partial\mathcal{M}_{4}}\varepsilon_{abc}\Big( \delta e^a\big( 2\psi (R^{bc}+4e^bk^c)\\\nonumber 
&
    -4 k^b \Diff\phi^c+4k^bk^c\varphi\big)\\\nonumber
&+\delta\varphi\big(2k^aR^{bc}+4k^ak^be^c\big)+\delta \phi^a\big(-4k^bT^c \big)\Big )\\\nonumber
    &+\delta\omega^{ab}\big(  \varepsilon_{abc}(2\psi T^c-2\phi^c e^dk_d-2k^c\diff\varphi)\\ &-4\varepsilon_{acd} k^c\phi^de_b\big)\Big).
\end{align}
On-shell, modified action is different from zero, and 
using the holographic dictionary, we have 
\begin{align}
\delta S_{\mathrm{mod}}=\delta W=\int_{\partial \mathcal{M}_{4}}\Big(&\delta e^a\tau_a+\frac{1}{2}\delta\omega^{ab}\sigma_{ab}\nonumber\\
&+\delta\varphi o_\varphi+\delta\phi^a o_{a}\Big),
\end{align}
where $W$ is the generating functional of connected Green's functions in the dual QFT.
From this, we can read out the one-point correlation functions. They are given by 
\begin{align}\nonumber
\tau_a=\langle \mathcal{T}_a\rangle_{\mathrm{QFT}}&=\frac{1}{16\pi G}\varepsilon_{abc}\big(2\psi (R^{bc}+4e^bk^c)\\
&\hspace{14mm}-4k^b\Diff \phi^c+4k^bk^c\varphi\big), \\ \nonumber
\sigma_{ab}=\langle\mathcal{S}_{ab}\rangle_{\mathrm{QFT}}&=\frac{1}{16\pi G}\varepsilon_{abc}(2\psi T^c 
-2\phi^c e^dk_d \\&\hspace{10mm}-2k^c\diff\varphi)-4\varepsilon_{acd} k^c\phi^de_b,\\
    o_\varphi=\langle \mathcal{O}_\varphi\rangle_{\mathrm{QFT}}&=\frac{1}{8\pi G}\varepsilon_{abc}(k^aR^{bc}+2k^ak^be^c),\\
o_a=\langle\mathcal{O}_a\rangle_\mathrm{QFT}&=-\frac{1}{4\pi G}\varepsilon_{abc}k^bT^c.
\end{align}
Fields $k^a$ and $\psi$ have to satisfy certain constraints that we present in the subsection \ref{geom}, but are not fixed by the boundary sources. These constraints are given either by obtaining the equations of motion for the bulk action in the radial direction or by dimensionally reducing the constraints found in \cite{Banados:2006fe}. They will be helpful in the discussion concerning the holographic Weyl anomaly. 

Alternatively, we can deform the boundary theory by adding the boundary term originating from the dimensional reduction of 5-dimensional CS gravity. This term is given by
\begin{equation}\label{csmod}
\frac{1}{64\pi G}\int_{\partial\mathcal{M}_4}\varepsilon_{ABCD}\Big( \frac{4}{3}\hat{e}^A\hat{e}^B\hat{e}^C\hat{\phi}^D+ 4\hat{e}^A\hat{R}^{BC}\hat{\phi}^D\Big).
\end{equation}
On-shell, this term is divergent, and therefore holographic renormalization is necessary. This means that the boundary is first moved to some finite $\rho=\varepsilon$, and counterterms are added to cancel the divergences. As we are not interested in the nature of those terms, they will not be presented here. Additionally, we have to make sure that the variational principle is satisfied. This is again done by adding a suitable GHY-like boundary term. If the finite boundary term originating from the 5-dimensional CS gravity contains some of the fields $k^a$ or $\psi$, adding it would not change the one-point functions of the dual operators, given that the suitable GHY-like term is also included. As all terms in (\ref{csmod}), upon unpacking, involve one of those fields, we conclude that the structure of the one-point correlation functions remains the same. The full GHY-like term is now
\begin{align} 
S_{\text{GHY}}=&\frac{1}{16\pi G}\int_{\partial\mathcal{M}_4}\varepsilon_{abc}\Big (-4k^ak^be^c\varphi+k^a(R^{bc}+2k^be^c)\varphi\nonumber\\
&+ e^a(R^{bc}+2k^be^c)\psi-2e^ak^b\Diff \phi ^c\Big).
\end{align}

\subsection{Semi-classical bulk geometries}\label{geom}

Having computed the general form of the one-point correlation functions, we will provide some examples of solutions of the bulk equations of motion, thus identifying some semi-classical geometries that one could use to learn more about the dual QFT. Inserting the asymptotic FG expansions (\ref{e})-(\ref{varfi})
into the equations of motion (\ref{jna1})-(\ref{jna4}) we obtain the constraints that have to be satisfied by the boundary fields. Note that the first two equations (\ref{jna1})-(\ref{jna2}) are identically satisfied, while the remaining two yield  
\begin{align}\label{raspisano1}
\varepsilon_{abc}\big(\Diff \phi^a-2k^a\phi-2k^a\big)\big(R^{bc}+4e^bk^c\big)&=0,\\ \nonumber 
\label{raspisano2}
\varepsilon_{abc}\big[  (\diff \varphi-e_d\phi^d)\Diff k^c-(\diff \psi-k_d\phi^d)T^c&\\+(\Diff \phi^c-2e^c\psi-2k^c\varphi)e^dk_d\big]&=0,\\
\nonumber\label{raspisano3}
\varepsilon_{abc}\big[(\diff \varphi-e_d\phi^d)(R^{bc}+4e^bk^c)&\\
+2(\Diff \phi^b-2k^b\varphi-2e^b\psi)T^c \big]&=0,\\
\nonumber \label{raspisano4}
\varepsilon_{abc}\big[(\diff \psi-k_d\phi^d)(R^{bc}+4e^bk^c)&\\+2(\Diff \phi^b-2k^b\varphi-2e^b\psi)\Diff k^c \big]&=0.
\end{align}
The most obvious solution to the bulk equations of motion is the AdS$_4$ spacetime with $\hat{R}^{AB}+\hat{e}^A\hat{e}^B=0$ and vanishing torsion. However, this case is peculiar due to the fact that the scalar fields are completely arbitrary. In particular, this is an example of a bulk geometry with vanishing torsion for which the one-point function of the spin-current in the dual QFT can be non-vanishing due to the presence of bulk scalars. If we set the scalars to zero, we are left with the pure AdS$_4$ that corresponds to the vacuum state $\ket{0}$ in the boundary theory and has a vanishing spin-current at the boundary. 

In order to relate the discussed model to the physics of spin systems \cite{Gallegos:2020otk}, one should place the boundary field theory at a finite temperature. This is usually done by placing a black hole in the bulk. We will analyse the black hole with a flat horizon discussed in \cite{Canfora:2007xs, Brihaye:2013vsa}. In the Schwarzschild form, the 5-dimensional metric for this black hole is given by 
\begin{equation}\label{BHflat}
\diff s^2=-(r^2-\mu)\diff t^2+\frac{1}{(r^2-\mu)}\diff r^2+ r^2\big(\diff x^2+ \diff y^2+\diff z^2\big).\nonumber
\end{equation}
Analogous to the 3-dimensional case discussed in \cite{Tetradis:2011jn}, we can rewrite this metric in the FG form as
\begin{align}\label{BH4FG}\nonumber
\diff s^2=\frac{\diff \rho^2}{4\rho^2}+\frac{1}{\rho}\Big( (1&+\frac{\mu}{2}\rho +\frac{\mu^2}{16}\rho^2)(\diff x^2+ \diff y^2+\diff z^2)\\&-(1-\frac{\mu}{2}\rho +\frac{\mu^2}{16}\rho^2)\diff t^2 \Big). 
\end{align}
This black hole can have a non-vanishing torsion. The corresponding solution of the action (\ref{ovde}) is obtained by performing dimensional reduction. The resulting black hole has a similar metric but necessarily vanishing torsion (see Appendix \ref{fggauge} for the discussion on why it is legitimate to apply our formalism to this black hole). We get 
\begin{align}\label{ravno}
e^a=\delta_\mu^a\diff x^\mu,\;\;\; \omega^{ab}=0,\\
    k^a=\epsilon\frac{\mu}{4}  \delta_\mu^a\diff x^\mu,\;\;\;\varphi=1,\\
    \psi=\frac{\mu}{4}, \;\;\;\phi^a=0,
\end{align}
where $\epsilon=\pm 1$, depending on the value of $a$; for $a=0$ we have $\epsilon=-1$, and otherwise for $a=2,3$.   
From (\ref{ravno}), it is clear that the boundary is flat. This is appealing, considering the possible condensed matter applications. Also, it is clear that fields $e^a$ and $k^a$ are independent, as only one of them is proportional to the parameter $\mu$ of the black hole solution.
The Hawking temperature of this black hole solution, which corresponds to the temperature of the dual QFT, is $\frac{\sqrt{\mu}}{2\pi}$. One can readily check that the constraint equations (\ref{raspisano1})-(\ref{raspisano4}) are satisfied. It is also easy to see that this solution has a vanishing one-point function for the spin-current and therefore is not useful when dealing with spin systems. The spacetime components of the one-point function of the stress-energy tensor are given by 
\begin{align}
    \langle \mathcal{T}_0\rangle_{\mathrm{QFT}}=&\frac{3\mu^2}{32\pi G}\diff x^2\diff x^3,\\
    \langle \mathcal{T}_2\rangle_{\mathrm{QFT}}=&\frac{\mu^2}{32\pi G}\diff x^0 \diff x^3,\\
   \langle \mathcal{T}_3\rangle_{\mathrm{QFT}}=&-\frac{\mu^2}{32\pi G}\diff x^0 \diff x^2,&
\end{align}
and the one-point functions for operators $\mathcal{O}_\varphi$ and $\mathcal{O}_a$ are 
\begin{align}
    \langle \mathcal{O}_\varphi \rangle_{\mathrm{QFT}}=&-\frac{\mu^2}{32\pi G}\diff x^0 \diff x^2 \diff x^3,\\
     \langle \mathcal{O}_a\rangle_{\mathrm{QFT}}=&0.
\end{align}
Holographic considerations of dilaton gravity theories, with an emphasis on hydrodynamics, can also be found in \cite{Gursoy:2007cb, Gursoy:2009jd}.

Having in mind our goal to study holographic properties of CTG, one can further introduce a scalar field $f$ that is coupled with this background, neglecting backreaction. If the scalar is minimally coupled, this means that the action is given by 
\begin{equation}
\frac{1}{2}\int \diff ^4 x \sqrt{-g} g^{\mu\nu}\partial_\mu f \partial_\nu f.
\end{equation}
By solving the Klein-Gordon equation for the field $f$, we can obtain the spectrum of quasi-normal modes of a given black hole. Quasinormal modes are in holographic relation to relaxation times in the thermal state of the dual field theory. This was done in \cite{Oliva:2010xn}, and it is important to note that this spectrum can be found exactly without relying on some approximate methods. However, in the spirit of dilaton gravity, one may consider a more general actions describing a scalar field in the form of 
\begin{equation}
    \frac{1}{2}\int \diff ^4 x\sqrt{-g} \; \hat{\varphi}^Ng^{\mu\nu}\partial_\mu f \partial_\nu f,
\end{equation}
where $N$ is some positive number. The case of $N=1$ corresponds to the dimensional reduction of the minimally coupled scalar field in five dimensions. Modified Klein-Gordon equation, in this case, is given by (we can safely use partial integration with covariant derivatives as the torsion for this geometry is zero)
\begin{equation}\label{knm}
    \Box f-m^2f+N(\partial_\mu \ln \hat{\varphi})\partial^\mu f=0.
\end{equation}
Using the FG expansion for the scalar field $\hat{\varphi}=\frac{1}{\sqrt{\rho}}\left(1+\frac{\mu}{4}\rho\right)$, it is not hard to check that equation (\ref{knm}), for $N\in \mathbb{N}$, corresponds to the equation of ($4+N$)-dimensional black hole discussed in \cite{Oliva:2010xn}, and therefore shares the same spectrum of quasinormal modes. It is also interesting to note that equation (\ref{knm}) can be solved for non-integer values; for example, when $N=\frac{1}{2}$ the solution is given in terms of hypergeometric functions.

Finally, there is a solution to the equations of motion with nonvanishing torsion but vanishing scalar fields. The vielbeins match those in the black hole solution (\ref{BHflat}), but the spin-connection is modified by the presence of the contorsion tensor $K^{23}=K(r)\diff r$. It is interesting to note that the exact profile of the function $K(r)$ is not determined by the equations of motion. 

\subsection{Generalised holographic Weyl anomaly}

A CFT has a vanishing trace of the stress-energy tensor. However, when coupled to a curved background, an anomaly may appear, and the expectation value of the trace of the stress-energy tensor can be nonzero. Note, however, that in the present case, the scalar fields ruin the conformal symmetry in the usual sense, similar to the case of AdS/CFT with non-conformal branes \cite{Kanitscheider:2009as}. Our theory has a generalised conformal structure (for this reason we insisted on calling the boundary theory QFT and not CFT), and the holographic Weyl anomaly vanishes,  
\begin{equation}\label{anomalija}
e^a\langle \mathcal{T}_a\rangle_{\mathrm{QFT}}+\varphi\langle \mathcal{O}_\varphi\rangle_{\mathrm{QFT}} =\mathcal{A}=0,
\end{equation}
where $\mathcal{A}$ is the anomaly.
Note that the form of the conformal Ward identity (\ref{anomalija}) follows from the fact that the scaling dimensions of the operators dual to $e^a$ and $\varphi$ are the same, while bulk fields $\omega^{ab}$ and $\phi^a$ have no divergent parts, as can be seen from their asymptotic expansions. To derive (\ref{anomalija}), we use the constraints (\ref{raspisano1})-(\ref{raspisano4}) following from the bulk equations of motion.
We have neglected the total derivative $\diff(4\kappa\varepsilon_{abc}\phi^aR^{bc})$, as it can be removed by a suitable redefinition of the current. This is consistent with the fact that in three dimensions, for CFT, there should be no Weyl anomaly \cite{Henningson:1998gx}. Our result is similar to the considerations in \cite{Taylor:2017dly}. It is interesting to note that, in the case of 5-dimensional CS gravity, the nature of the holographic Weyl anomaly led authors to conclude that the boundary theory is a non-unitary CFT \cite{Banados:2004zt}. We have no reason to claim anything similar based on the derived result (\ref{anomalija}), as it is consistent with the expectations.

\section{Deformed 2D BF model and stress-energy tensor}\label{JT}

The previous procedure can be applied in the case of two spacetime dimensions. The action is given by \cite{Isler:1989hq, Chamseddine:1989wn}
\begin{equation}
\kappa\int_{\mathcal{M}_2}\varepsilon_{\hat{A}\hat{B}\hat{C}}\hat{\Phi}^{\hat{A}}\hat{F}^{\hat{B}\hat{C}}.
\end{equation}
For the $SO(2,1)$ gauge group, and the usual decomposition of the connection components, equations of motion imply vanishing torsion. Therefore, on-shell, this action is equivalent to the JT gravity,
\begin{equation}\label{JTeq}
    \frac{1}{16\pi G}\int_{\mathcal{M}_2} \diff^2x\; \varphi\big(R-2\Lambda\big).
\end{equation}
In the second-order formalism, this action has to be accompanied by a GHY term, given by $\frac{1}{8\pi G}\int_{\partial \mathcal{M}_2}\varphi K,$ where $K$ is the trace of the extrinsic curvature. Through holography, JT gravity is closely related to the SYK model \cite{Sachdev:2010um, Maldacena:2016hyu}. 

On-shell variation of the action (\ref{JTeq}) is given by 
\begin{equation}
    \kappa \int_{\partial\mathcal{M}_{2}}\varepsilon_{AB}\big(\hat{\varphi}\delta\hat{\omega}^{AB}+2\hat{\phi}^A\delta\hat{e}^B\big ).
\end{equation}
We can now plug in the expansions (\ref{e})-(\ref{varfi}), and extract the finite piece (part of the variation that does not contain $\rho$). As anticipated, terms that do contain powers of $\rho$ cancel exactly. The result is 
\begin{equation}
4\kappa\int_{\partial\mathcal{M}_{2}}( -\varphi\delta k+\psi\delta e).
\end{equation}
At this point, we will follow the same logic as in the 4-dimensional theory and add a boundary that will move the variation from the $k$ field to the scalar field $\varphi$. The boundary term is $4\kappa\int_{\partial\mathcal{M}_{2}}\varphi k$ (it clearly resembles the GHY term for JT gravity), and thus we obtain the following variation
\begin{equation}
\delta W=4\kappa\int_{\partial\mathcal{M}_{2}}(k\delta\varphi +\psi \delta e).
\end{equation} 
Note that our choice of boundary conditions is in the spirit of the "JT-like" boundary conditions from \cite{Ebert:2022ehb}, and therefore, does not give rise to the boundary Schwartzian dynamics - a low-energy limit of the SYK model. Our boundary conditions, on the other hand, are in perfect analogy with the type of boundary conditions discussed in \cite{Banados:2006fe, Cvetkovic:2017fxa}, which are useful for holographic considerations. 

The one-point functions are 
\begin{align}\label{2dT}
\langle \mathcal{T} \rangle=4\kappa \psi,\\
\langle \mathcal{O}_\varphi \rangle=4\kappa k.
\end{align}
Constraints that have to be satisfied due to the bulk equations of motion are 
\begin{align}\label{jne2d1}
    \diff \varphi=e\phi, \hspace{5mm} \diff \psi=k\phi,\\\label{jne2d2}
    \diff \phi=2k\varphi +2e\psi.
\end{align}
Note that $e \langle \mathcal{T} \rangle+\varphi \langle \mathcal{O}_\varphi \rangle=0$, up to boundary terms, confirming there is no Weyl anomaly.
Boundary theory is 1-dimensional QFT, which is just ordinary quantum mechanics. The one-point function $\langle \mathcal{T}\rangle$, therefore, corresponds to the expectation value of the Hamiltonian.

This theory has a black hole solution similar to the one discussed in section  (\ref{geom}), obtained by dimensional reduction of a spinless BTZ black hole, 
\begin{equation}
    \diff s^2=-(r^2-\mu)\diff t^2+\frac{1}{r^2-\mu}\diff r^2, \hspace{5mm} \hat{\varphi}=r.
\end{equation}
In the FG gauge, we have the following expressions: $e=\diff t$, $k=-\frac{\mu}{4}\diff t$, $\varphi=1$, $\psi=\frac{\mu}{4}$ and $\phi=0$. It is easy to check that equations (\ref{jne2d1}) and (\ref{jne2d2}) are indeed satisfied. Moreover, we can use the one-point function (\ref{2dT}) to obtain the thermodynamics of this black hole solution. In the case of three dimensions and Einstein-Hilbert gravity, this was done in \cite{Tetradis:2011jn}, and for JT gravity, entropy was computed in \cite{Almheiri:2014cka}. We first note that avoiding conical singularity in the Euclidean signature results in the temperature $T=\frac{\sqrt{\mu}}{2\pi}$, as before. Furthermore, we have  
\begin{equation}
 \frac{\delta W}{\delta e^t_t}=\frac{\delta W}{\delta g^{tt}}\frac{\partial( e^t_t e^{tt})}{\partial e^t_t}=-2\frac{\delta W}{\delta g^{tt}}=\langle \mathcal{T}\rangle =E,
\end{equation}
and thus from (\ref{2dT}) follows that the energy of the black hole is $\kappa\mu$. If we write $\kappa=\frac{1}{16\pi G}$, we get \begin{equation}
    E=\frac{\mu}{16\pi G}.
\end{equation}
Using the relation $T\diff S=\diff E$, together with the fact that entropy is zero for zero temperature, we obtain 
\begin{equation}
    S=\frac{\sqrt{\mu}}{4G}.
\end{equation}
This result coincides with the black hole entropy calculated in the metric formulation of JT gravity \cite{Witten:2020ert, Lemos:1996bq}.

Finally, we can deform the theory by adding a boundary term originating from the CS gravity action in three dimensions,
\begin{equation}
2\kappa \int_{\partial \mathcal{M}_2}\varepsilon_{AB}\phi^A e^B.
\end{equation}
Total action is now on-shell divergent, and the machinery of holographic renormalization has to be used. Putting the boundary at some finite $\rho=\varepsilon$, we add a counterterm 
\begin{equation}
    \frac{2}{\varepsilon}\kappa \int_{\mathcal{M}_2}\varphi e, 
\end{equation}
that removes the divergences. In addition, we have to modify the previously added GHY term, so that we still have well-defined boundary conditions. It is easy to check that, upon adding the relevant GHY term, the one-point functions in the dual theory remain the same.
The total GHY term, in this case, is given by 
\begin{equation}
   2\kappa \int_{\partial \mathcal{M}_2}( \psi e+\varphi k ).
\end{equation}

\section{Gravitational Wilson lines}\label{wlline}

It was argued in a series of papers that in the case of 3-dimensional CS gravity, the bulk Wilson line observable corresponds to a bi-local operator in the dual QFT. The same is expected to be true in the 2-dimensional model of topological gravity (see, for example, \cite{Ammon:2013hba, Ebert:2022ehb, Iliesiu:2019xuh, DHoker:2019clx, Bhatta:2016hpz}). In the context of gauge theories of gravity, Wilson lines are closely related to heavy particles moving in the gravitational field \cite{Freidel:2006hv}. We would like to understand the importance of Wilson lines in the holographic setting. Through this section, we will not threat dilaton fields and therefore our analysis is applicable (and possibly better suited) for other gauge theories of gravity. We are not a priori claiming that this object will be a two-point correlation function in the boundary QFT, but as we will see, in some cases, this might be true. We start with the particle action 
\begin{equation}
    S_{\mathrm{par}}=-\int \mathrm{Tr}(KA^h_\tau)\diff \tau,
\end{equation}
where the trace is given in the explicit representation of the $SO(3,2)$ algebra using gamma matrices. $K$ is a fixed algebra element, given by $K=mP_0+ 
\frac{1}{2}sJ_{23}$. Field $h$ is a Lorentz-algebra-valued one-form, and is used as a gauge parameter to gauge transform field $A_\tau= A_\mu \frac{\diff x^\mu}{\diff \tau}$. As explained in \cite{Freidel:2006hv}, this can be interpreted as a gravitational Wilson line insertion in the bulk. The Wilson line depends on the choice of representation. We work in an infinite dimensional irreducible representation labelled by two numbers $(m,s)$, representing the particle's mass and spin. 
\begin{figure}[h!]
    \centering
     \begin{tikzpicture}[scale=0.65,line cap=round,line join=round,x=1.0cm,y=1.0cm]
\fill[line width=2.pt,color=bcduew,fill=bcduew,fill opacity=0.20000000298023224] (1.,2.8) -- (1.,-3.) -- (4.8,-2.3) -- (4.8,3.51294) -- cycle;
\draw [line width=1.pt] (1.,2.8)-- (1.,-3.);
\draw [line width=1.pt] (1.,-3.)-- (4.8,-2.3);
\draw [line width=1.pt] (4.8,-2.3)-- (4.8,3.51294);
\draw [line width=1.pt] (4.8,3.51294)-- (1.,2.8);
\draw [color=red] (1.5,-0.5) .. controls   (4.3492, -1.81032) and (6.29815, -1.0928) .. (4.5,0.);
\begin{scriptsize}
\draw [fill=red] (1.5,-0.5) circle (2.5pt);
\draw [fill=red] (4.5,0.) circle (2.5pt);
\end{scriptsize}
\end{tikzpicture}
    \caption{Probe particle world-line, interpreted as a gravitational Wilson line in the bulk.}
\end{figure}
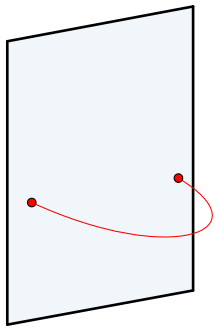

The (Euclidean) path integral is given by 
    \begin{equation}\label{patintegral}
        \int \mathcal{D}\mathcal{P}\mathcal{D}K\mathcal{D}h \;e^{-\int \diff \tau \big(-\mathrm{Tr}(KA_\tau^h)+\mathrm{L.M.}  \big)}.
    \end{equation}
The action terms denoted by $\mathrm{L.M.}$ are the constrain terms fixing the two Casimirs of the $SO(3,2)$ algebra,
\begin{align}\nonumber
        \mathrm{L.M.}=\lambda_1\Big( \frac{1}{2}J^{\hat{A}\hat{B}}J_{\hat{A}\hat{B}}-c_2 \Big)\\+\lambda_2\Big( \frac{1}{16}\varepsilon_{\hat{A}\hat{B}\hat{C}\hat{D}}\varepsilon^{\hat{E}\hat{F}\hat{G}\hat{H}}J^{\hat{A}\hat{B}}J^{\hat{C}\hat{D}}J_{\hat{E}\hat{F}}J_{\hat{G}\hat{F}}-c_4 \Big),
\end{align}
where $\lambda_{1}$ and $\lambda_{2}$ stand for the Lagrange multipliers, and generators $J^{\hat{A}\hat{B}}$ are defined in \cite{Freidel:2006hv}. 
   
To obtain the two-point function on the boundary, it was necessary to include the integration over paths ($\mathcal{P}$) connecting the two given boundary points because the bulk gauge curvature is not zero. If we were to consider the topological BF action, then there would be no need to include this integration and one could assume that the field $h$ is valued in the whole $SO(3,2)$ algebra, not only in the Lorentz subalgebra. In the latter case, the interpretation of particles as Wilson lines is exact, as explained in \cite{Freidel:2006hv}. 

In \cite{Ammon:2013hba}, it was important to restrict to a concrete representation of a gauge group in order to make the connection with the entanglement entropy. In four spacetime dimensions, it is not expected that Wilson lines should be able to reproduce boundary entanglement entropy, but it is nevertheless possible to study extended operators in the dual field theory. One can formulate surface defects (strings) in the bulk, that anchor the boundary along a given curve \cite{Fairbairn:2008hy}, that might be related to the entanglement entropy, but we will not pursue this question here.
What we are discussing in this section is similar to the construction of the two-point function from \cite{Erdmenger:2017pfh}.
 
Quadratic Casimir $c_2\sim m^2$ should be assumed large so that we can safely use the saddle point approximation of the path integral (\ref{patintegral}). Using the standard holographic dictionary, a scalar bulk field of mass $m$ is dual to the primary operator of scaling dimension $\Delta=\frac{D}{2}+\sqrt{\frac{D^2}{4}+m^2}$. In the limit of a large mass, the scaling dimension coincides with the $m\sim\sqrt{c_2}$. We will also assume that the spin of the particle is either zero, or large, in order to apply the semi-classical approximation safely. However, we will work in a regime where $s\ll m$, so that the Casimir operators of the $SO(3,2)$ group, $c_2$ and $c_4$, reproduce the well-known Casimir operators in the Minkowski spacetime (where the concept of a particle is well-defined). Of course, the difference between the entanglement entropy and our two-point correlation function is that the entanglement entropy is UV divergent in field theory, while the two-point function has to be renormalized. The saddle point approximation yields  
\begin{equation}
        \langle \mathcal{O}(x_1)\mathcal{O}(x_2)\rangle=\lim_{\varepsilon\rightarrow 0}\varepsilon^{2m}e^{-S_{\mathrm{par|_{on-shell}}}}.
\end{equation}
One can then calculate, using the described prescription, the two-point function in different states with semi-classical bulk. For example, let us discuss the case of AdS${_\mathrm{4}}$ bulk spacetime and a spinless particle. Equation of motion obtained by varying $h$ is \cite{Freidel:2006hv} 
\begin{equation}
    \frac{\diff z^\mu}{\diff \tau}p^\nu=\frac{\diff z ^\nu}{\diff \tau} p^\mu,
\end{equation}
which is solved by $p^\mu= m\frac{\diff z^\mu}{\diff \tau}$ ($\tau$ being the affine parameter of a geodesic). The equation following the variation with respect to the path $\mathcal{P}$ then reproduces the standard geodesic equation. For simplicity, we use coordinate $z=\sqrt{\rho}$, such that the metric of AdS${_\mathrm{4}}$ reads
\begin{equation}
    \diff s^2=\frac{\diff z^2-\diff t^2+\diff x^2+\diff y^2}{z^2}.
\end{equation}
Focusing on constant time correlation functions between two boundary points at distance $L$ (this distance is defined with respect to the flat boundary metric), geodesics are given by semi-circles, connecting two boundary points \cite{geod}. We can then calculate the on-shell action and obtain the final result for the two-point correlation function. The result is given by
\begin{equation}
     \langle \mathcal{O}(x_1)\mathcal{O}(x_2)\rangle\sim \frac{1}{L^{2m}}.
\end{equation} 
Finally, consider a pair of QFT states, $\ket{\psi_{g_1}}$ and $\ket{\psi_{g_2}}$, that are dual to a pair of very distinct semi-classical bulk geometries, $g_{1}$ and $g_{2}$, respectively.
One could ask what should be the dual description of a superposition of these two QFT states, e.g. $\frac{1}{\sqrt{2}}(\ket{\psi_{g_1}}+\ket{\psi_{g_2}})$. Based on the considerations of the entanglement entropy in \cite{Almheiri:2016blp, Papadodimas:2015jra} we believe that the answer to this question would be that it is a weighted sum of correlation functions corresponding to the states $\ket{\psi_{g_1}}$ and $\ket{\psi_{g_2}}$.

\section{Conclusion and outlook}\label{kraj}

The holographic analysis of the Chamseddine's even-dimensional topological gravity is presented for all even dimensions. One-point 
correlation functions of the dual QFT are obtained and the generalised holographic Weyl anomaly is discussed. A method of computing two-point correlation functions in terms of gravitational Wilson lines is also proposed. 
We emphasized the role of boundary GHY-like terms in defining the bulk action with appropriate holographic interpretation. In that respect, our results contribute to a better understanding of holography for Riemann-Cartan spaces \cite{Erdmenger:2023hne}. We should also point out the similarity of the CTG action (\ref{ovde}) to that of Brans-Dicke theory (scalar-tensor modified gravity theory) \cite{Inostroza:2014vua}. This theory can be considered as a particular frame change of $f(R)$ models. Some (but not complete) progress has been made to understand holography in $f(R)$ models, especially in three dimensions \cite{Loran:2013fca}.

Although we started from a concrete gravity theory in the bulk and aimed at finding its holographic features, our motivation largely came from the considerations regarding the holographic description of 4-dimensional hydrodynamics of spin-systems \cite{Gallegos:2020otk} (see also \cite{Hashimoto:2013bna}). The even-dimensional topological theory of gravity that we focused on also predicts the possibility of obtaining the nonzero spin-current for the boundary QFT, but now an odd-dimensional one. This is due to the first-order treatment of our gravity theory. In particular, it would be impossible to formulate the CTG action (\ref{ovde}) with $SO(3,2)$ gauge symmetry, assuming zero torsion. However, it is interesting to note that the boundary spin-current is nonvanishing even for some torsion-free bulk geometries, which can be relevant for the physics of spin systems. A potential further investigation could involve suitable modelling of the bulk scalar fields to capture some interesting features of the dual spin systems. 

{\bf Acknowledgments}\newline
We thank Olivera Miskovic and Rodrigo Olea for useful discussions on similar topics, and to Dusan Novicic for useful comments on boundary conditions. We also thank C. Brukner, A.C. de la Hamette and V. S. Kabel for the discussion on the superposition of semi-classical geometries. 
Work of D.D. and D.G. is supported by the funding provided by the Faculty of Physics, University of Belgrade, through grant number 451-03-47/2023-01/200162 by the Ministry of Science, Technological Development and Innovations of the Republic of Serbia.

\appendix
\setcounter{figure}{0}
\renewcommand{\thefigure}{A\arabic{figure}}
\renewcommand{\theequation}{\Alph{section}.\arabic{equation}}

\section*{APPENDIX}

\section{Some basic algebra behind the Chern-Simons theory of gravity}

We start with an observation that, up to a boundary term, 5-dimensional CS gravity action (\ref{CS5D}) is equivalent to the CS gauge theory action,  
\begin{align}
\label{CSgauge}
-\frac{ik}{3}\int_{\mathcal{M}_{5}}\tr\left( \hat{\mathcal{F}}^{2}\hat{\mathcal{A}}-\frac{1}{2}\hat{\mathcal{F}}\hat{\mathcal{A}}^{3}+\frac{1}{10}\hat{\mathcal{A}}^{5}\right),
\end{align}
for the conformal gauge group $SO(4,2)$. The corresponding gauge connection and field strength are
\begin{align}\label{AF}
\hat{\mathcal{A}}&=\frac{1}{2}\hat{\Omega}^{\mathbb{A}\mathbb{B}}J_{\mathbb{A}\mathbb{B}}+\hat{E}^{\mathbb{A}}J_{\mathbb{A}5}, \nonumber\\
\hat{\mathcal{F}}&=\frac{1}{2}(\hat{R}^{\mathbb{A}\mathbb{B}}+\hat{E}^{\mathbb{A}}\hat{E}^{\mathbb{B}})J_{\mathbb{A}\mathbb{B}}+\hat{T}^{\mathbb{A}}J_{\mathbb{A}5},
\end{align}
where $J_{\mathbb{A}\mathbb{B}}$ and $J_{\mathbb{A}5}$ are $SO(4,2)$ 
generators. The gauge group index $\mathbb{A}$ takes values from $0$ to $4$; we decompose it as $\mathbb{A}=(A,4)$ where $A=0,1,2,3$ is the standard $SO(3,1)$ Lorentz index. Group $SO(4,2)$ acts in the 6-dimensional flat space with signature $(-++++-)$ as the isometry group of AdS$_{5}$. The generators of $SO(4,2)$ satisfy the following commutation relations:
\begin{align}
[J_{\mathbb{A}\mathbb{B}},J_{\mathbb{C}\mathbb{D}}]&=G_{\mathbb{A}\mathbb{D}}J_{\mathbb{B}\mathbb{C}}+G_{\mathbb{B}\mathbb{C}}J_{\mathbb{A}\mathbb{D}}-(\mathbb{A}\leftrightarrow \mathbb{B}),  \\
[J_{\mathbb{A}\mathbb{B}},J_{\mathbb{C}5}]&=G_{\mathbb{B}\mathbb{C}}J_{\mathbb{A}5}-G_{\mathbb{A}\mathbb{C}}J_{\mathbb{B}5},\\
[J_{\mathbb{A}5},J_{\mathbb{C}5}]&=J_{\mathbb{A}\mathbb{C}},\end{align}
where $G_{\mathbb{A}\mathbb{B}}=(-++++)$. A representation of the algebra is provided by 5-dimensional gamma matrices $\Gamma_{\mathbb{A}}$ satisfying the Clifford algebra $\{\Gamma_{\mathbb{A}},\Gamma_{\mathbb{B}}\}=2G_{\mathbb{A}\mathbb{B}}\mathbb{1}_{4\times 4}$, 
\begin{align}
J_{\mathbb{A}\mathbb{B}}&=\frac{1}{2}\Gamma_{\mathbb{A}\mathbb{B}}=\frac{1}{4}[\Gamma_{\mathbb{A}},\Gamma_{\mathbb{B}}],\\
J_{\mathbb{A}5}&=\frac{1}{2}\Gamma_{\mathbb{A}}. 
\end{align}
Finally, we can use ordinary 4-dimensional gamma matrices, satisfying $\{\gamma_{A},\gamma_{B}\}=-2G_{AB}\mathbb{1}_{4\times 4}$ with $G_{AB}=(-+++)$, to define the 5-dimensional ones as $\Gamma_{\mathbb{A}}=(-i\gamma_{A},\gamma_{5})$. Also, $\Gamma_{4}=\gamma_{5}=-i\gamma_{0}\gamma_{1}\gamma_{2}\gamma_{3}$ and $\gamma_{5}^{2}=\mathbb{1}_{4\times 4}$. Furthermore, we have  $\Gamma_{AB}=-\frac{1}{2}[\gamma_{A},\gamma_{B}]=i\sigma_{AB}$ and $\Gamma_{A4}=\frac{1}{2}[\Gamma_{A},\Gamma_{4}]=-\frac{i}{2}[\gamma_{A},\gamma_{5}]=-i\gamma_{A}\gamma_{5}$.
In this representation, we have the following trace identities for the generators: 
\begin{align}
\tr(J_{\mathbb{A}\mathbb{B}}J_{\mathbb{C}\mathbb{D}}J_{\mathbb{E}5})&=\frac{i}{2}\varepsilon_{\mathbb{A}\mathbb{B}\mathbb{C}\mathbb{D}\mathbb{E}},\nonumber\\
\tr(J_{\mathbb{A}\mathbb{B}}J_{\mathbb{C}5}J_{\mathbb{D}5})&=\frac{1}{2}(G_{\mathbb{A}\mathbb{B}}G_{\mathbb{C}\mathbb{D}}-G_{\mathbb{A}\mathbb{C}}G_{\mathbb{B}\mathbb{D}}+G_{\mathbb{B}\mathbb{C}}G_{\mathbb{A}\mathbb{D}}),\nonumber\\
\tr(J_{\mathbb{A}5}J_{\mathbb{C}5}J_{\mathbb{D}5})&=0.
\end{align}
The generators can be further decomposed 
into $J_{AB}$, $J_{A4}$, $J_{A5}$ and $J_{45}$. The subgroup $SO(3,2)$ (the isometry group of AdS$_\mathrm{4}$) is generated by $J_{AB}$ and $J_{A5}$, and the corresponding gauge connection and field strength are
\begin{align}
\hat{\Omega}&=\frac{1}{2}\hat{\Omega}^{\hat{A}\hat{B}}J_{\hat{A}\hat{B}}=\frac{1}{2}\hat{\omega}^{AB}J_{AB}+\hat{e}^{A}J_{A5},\\
\hat{F}&=\mathrm{d}\hat{\Omega}+\hat{\Omega}^{2}=\frac{1}{2}\hat{F}^{AB}J_{AB}+\hat{T}^{A}J_{A5}.
\end{align}
The $SO(3,2)$ gauge group index $\hat{A}$ is decomposed as $\hat{A}=(A,5)$ where again $A=0,1,2,3$. 

The starting point of the dimensional reduction is to decompose the 5-dimensional vielbein $\hat{E}^{\mathbb{A}}$ and spin-connection $\hat{\Omega}^{\mathbb{A}\mathbb{B}}$ into components and make the following identification: $\hat{E}_\mu^A=\hat{e}_\mu^A$, $\hat{\Omega}_\mu^{AB}=\hat{\omega}_\mu^{AB}$, $\hat{\Omega}_4^{4A}=\hat{\phi}^A$ and $\hat{E}_4^4=\hat{\varphi}$. The spacetime indices $\mu,\nu$ take values from $0$ to $3$. All other components are truncated, which is consistent with the residual $SO(3,2)$ gauge symmetry.
The scalar fields can be combined into a single object, 
\begin{align}
\hat{\Phi}&=\hat{\Phi}^{\hat{A}}J_{4\hat{A}}=\hat{\Phi}^{A}J_{4A}+\hat{\Phi}^{5}J_{45}\nonumber\\
&\equiv\hat{\phi}^{A}J_{4A}+\hat{\varphi}J_{45}=\frac{i}{2}\hat{\phi}^{A}\gamma_{A}\gamma_{5}+\frac{1}{2}\hat{\varphi}\gamma_{5}.
\end{align}
Performing dimensional reduction of the 5-dimensional CS gravity action (\ref{CS5D}), one obtains, up to a boundary term, the 4-dimensional CTG action (\ref{ovde}) that can be compactly written as
\begin{equation}
S_{\text{CTG}}\propto \int_{\mathcal{M}_{4}} \tr(\hat{\Phi}\hat{F}\hat{F}),     
\end{equation}
which is manifestly invariant under $SO(3,2)$ gauge transformations because both $\hat{\Phi}$ and $\hat{F}$ change by a commutator, e.g. $\delta_{\epsilon}\hat{\Phi}=[\epsilon,\hat{\Phi}]$, where $\epsilon$ is an $SO(3,2)$ gauge parameter. 

We define $\varepsilon_{0123}=+1$ and $\varepsilon^{0123}=-1$. Also, we have
\begin{align}
\varepsilon^{ABCD}\varepsilon_{ABEF}&=-2!(\delta^{C}_{E}\delta^{D}_{F}-\delta^{C}_{F}\delta^{D}_{E})\\
\varepsilon^{ABCD}\varepsilon_{ABCE}&=-3!\delta^{D}_{E}\\
\varepsilon^{ABCD}\varepsilon_{ABCD}&=-4!
\end{align}

\section{Fefferman-Graham gauge}\label{fggauge}

The fact that we can choose the FG gauge can also be shown directly, without relying on dimensional reduction. First, we note that in the general AdS gauge invariant gravity theory, we can use the AdS gauge symmetry to set $\hat{\omega}_\rho^{AB}=0$, and $\hat{e}_\rho=-\frac{\diff \rho}{2\rho}$ \cite{Cvetkovic:2017fxa}. Further, we can use the diffeomorphism invariance. Note that there are only $D-1$ independent diffeomorphisms, in accordance with the CS case. To do so, it is important to note that the generic solution of equation (\ref{istakaocs}) satisfies \cite{Mignemi:1998us} $$\begin{pmatrix}
\hat{F}_{i\rho}^{\hat{A}\hat{B}}\\
\Diff_\rho\hat{\Phi}^{\hat{C}}
\end{pmatrix}=\mathcal{N}^j \begin{pmatrix}
\hat{F}_{ji}^{\hat{A}\hat{B}}\\
\Diff_j\hat{\Phi}^{\hat{C}}
\end{pmatrix},
$$ where parameters $\mathcal{N}^j$ are connected with diffeomorphism, and we can use the diffeomorphism symmetry to put $\mathcal{N}^i=0$ \cite{Mignemi:1998us, Banados:2006fe}.
Condition $\hat{F}_{\rho i}=0$ is the same one as in the CS case, and therefore we get the same expansion of the vielbein and the spin-connection. In order to deal with the scalar fields, for convenience, we will use a faithful representation of the $SO(2,3)$ algebra, given by gamma matrices. In this case, we can write the scalar field as $\hat{\Phi}=i\hat{\phi}^A\frac{\gamma_A\gamma_5}{2}+\hat{\varphi}\frac{\gamma_5}{2}$. Gauge transformations of the field $\hat{\Phi}$, in this representation, are generated by a commutator, and the commutator enters the definition of the covariant derivative. Therefore, we can solve $$\Diff_\rho\hat{\Phi}^{\hat{C}}=0,$$
in a similar way one solves the equation $\hat{F}_{\rho i}=0$, 
and derive the form of the FG expansion. This result is the same as the one obtained from the dimensional reduction. 
It is important to note that our holographic ansatz works for a generic solution of the bulk equations, but cannot be used for an arbitrary solution. We, therefore take the philosophy that we care only about those spacetimes with the given asymptotic form of bulk fields, similar to \cite{Blagojevic:2013bu}, hoping that this form will incorporate all the interesting holographic properties of our model. As we have shown in the paper, the black hole solution, although not of a generic type, can be put in the desired FG gauge. Even if the full consideration would introduce a more general FG expansion, we see that all other terms, in this case, must be zero, and we can safely use the computed one-point functions.
Also, note that in analogy with the consideration in \cite{Gallegos:2020otk}, one can choose a more general gauge choice in order to describe more general black holes. Dimensionally reducing those solutions found in \cite{Gallegos:2020otk} one may hope to acquire a better understanding of spin systems 3-dimensional spacetime.

\section{Generalisation to $D=2n$ dimensional bulk}

The previous discussion can be straightforwardly lifted to an arbitrary even-dimensional spacetime. First, the FG gauge can be obtained in the same manner as before, resulting in the same form of the vielbein, spin-connection and scalar fields. It is interesting that the FG gauge has the same form in different dimensions. Note, however, that the form of the CTG action is technically different in different dimensions, and this is in contrast with the Einstein-Hilbert case where the action is linear in curvature in any number of dimensions. The 2n-dimensional CTG action is given by 
\begin{equation}
\int_{\mathcal{M}_{2n}}\varepsilon_{CA_1B_1\dots A_nB_n} \hat{\Phi}^C\hat{F}^{A_1B_1}\dots \hat{F}^{A_nB_n}\equiv  \int_{\mathcal{M}_{2n}}\varepsilon \hat{\Phi}\hat{F}^n,
\end{equation}
where we have suppressed indices for the sake of simplicity.
Variation of this action yields 
\begin{align}\nonumber
    &\delta S^{(2n)}_{CTG}\vert_{\mathrm{on-shell}}=\kappa\int_{\mathcal{M}_{2n}} \varepsilon\Big[\delta e \Big( 4n\psi(R+4ek)^{n-1}\\\nonumber&+8n(n-1)\varepsilon \phi \Diff k (R+4ek)^{n-2} \Big)\\\nonumber&+\delta k \Big( -4 n\varphi(R+4ek)^{n-1}-8 n(n-1)\phi T(R+4ek)^{n-2}  \Big) \\\nonumber&\delta \omega \Big( - 4n(n-1)\varphi \Diff k (R+4ek)^{n-2}\\\nonumber&+4n(n-1) \psi T(R+4ek)^{n-2}\\\nonumber & -4n(n-1)\phi e^a k_a (R+4ek)^{n-2}\\& +8n(n-1)(n-2)\phi\Diff k T(R+4ek)^{n-3} \Big) \Big].
\end{align}
We used the FG expansion and focused on the finite terms, as before. To have a well-defined variation principle, we deform the theory by adding a surface GHY-like term, 
\begin{align}\nonumber
&S^{(2n)}_{GHY}=4\kappa\int_{\mathcal{M}_{2n}}n\sum_{j=0}^{n-1}\frac{1}{n-j}\varepsilon k \varphi \binom{n-1}{j}R^j(4ek)^{n-1-j}\\&+2n(n-1)\sum_{j=0}^{n-2}\frac{1}{n-1-j}k\phi T \binom{n-2}{j}R^j(4ek)^{n-2-j}. 
\end{align}
From the variation of the total action (including $S^{(2n)}_{GHY}$), we can read off the one-point functions as before. The result is 
\begin{align}\nonumber
    &\tau_a=\langle \mathcal{T}_a\rangle_{QFT}=\kappa\varepsilon \Big(  4n\psi(R+4ek)^{n-1}\\\nonumber &+8n(n-1) \phi \Diff k (R+4ek)^{n-2}\\\nonumber&+\sum_{j=0}^{n-1}\frac{4n}{n-j} k\varphi\binom{n-1}{j}(n-1-j)R^je^{n-2-j}(4k)^{n-1-j}\\\nonumber 
    &+\sum_{j=0}^{n-2}\frac{8n(n-1)}{n-1-j}\binom{n-2}{j}\Diff(k\phi R^j (4ek)^{n-2-j})\\
    &-\sum_{j=0}^{n-2}\phi T\frac{n-2-j}{n-1-j}\binom{n-2}{j}R^j(4k)^{n-2-j}e^{n-3-j},
    \end{align}

    \begin{align}
    \nonumber
     &\sigma_{ab}=\langle \mathcal{S}_{ab}\rangle_{QFT}=\kappa\Big[ \varepsilon\Big(- 4n(n-1)\varphi \Diff k (R+4ek)^{n-2}\\\nonumber&+4n(n-1) \psi T(R+4ek)^{n-2} \nonumber\\
     &-4n(n-1)\phi e^a k_a (R+4ek)^{n-2}\\\nonumber& +8n(n-1)(n-2)\phi\Diff k T(R+4ek)^{n-3}\times\\\nonumber &\times4n \sum_{j=0}^{n-1}\frac{1}{n-j}\binom{n-1}{j}j\Diff(k\varphi R^{j-1}(4ek)^{n-1-j})
    \\\nonumber&+8n(n-1)\sum_{j=0}^{n-2}\frac{1}{n-1-j}\binom{n-2}{j}j\Diff(k\phi T(4ek)^{n-2-j}) \Big)\\&-8n(n-1)\varepsilon\sum_{j=0}^{n-2}\frac{1}{n-1-j}\binom{n-2}{j}k\phi e R^j(4ek)^{n-2-j}  \Big].\nonumber
    \end{align}

    \begin{align}
     o_\varphi&=\langle \mathcal{O}_\varphi \rangle_{QFT}\nonumber\\
     &=\kappa\varepsilon\Big( 4n\sum_{j=0}^{n-1}\frac{1}{n-j}\binom{n-1}{j}kR^{i}(4ek)^{n-1-j}\Big).
     \end{align}

     \begin{align}
     o_a&=\langle \mathcal{O}_a\rangle_{QFT}\\
     &=\kappa \varepsilon\Big(-8n(n-1)\sum_{j=0}^{n-2}\frac{1}{n-1-j}\binom{n-2}{j}kTR^{j}4ek^{n-2-j}   \Big). \nonumber
\end{align}


\begin{thebibliography}{99}

\bibitem{Maldacena:1997re}
J.~M.~Maldacena,
\emph{``The Large N limit of superconformal field theories and supergravity,''}
Adv. Theor. Math. Phys. \textbf{2} (1998), 231-252
doi:10.1023/A:1026654312961

\bibitem{Witten:1998qj}
E.~Witten,
\emph{``Anti-de Sitter space and holography,''}
Adv. Theor. Math. Phys. \textbf{2} (1998), 253-291
doi:10.4310/ATMP.1998.v2.n2.a2

\bibitem{Engelsoy:2016xyb}
J.~Engels\"oy, T.~G.~Mertens and H.~Verlinde,
\emph{``An investigation of AdS$_{2}$ backreaction and holography,''}
JHEP \textbf{07} (2016), 139
doi:10.1007/JHEP07(2016)139

\bibitem{Sachdev:2010um}
S.~Sachdev,
\emph{``Holographic metals and the fractionalized Fermi liquid,''}
Phys. Rev. Lett. \textbf{105} (2010), 151602
doi:10.1103/PhysRevLett.105.151602

\bibitem{Maldacena:2016hyu}
J.~Maldacena and D.~Stanford,
\emph{``Remarks on the Sachdev-Ye-Kitaev model,''}
Phys. Rev. D \textbf{94} (2016) no.10, 106002
doi:10.1103/PhysRevD.94.106002

\bibitem{Banados:2006fe}
M.~Banados, O.~Miskovic and S.~Theisen,
\emph{``Holographic currents in first order gravity and finite Fefferman-Graham expansions,''}
JHEP \textbf{06} (2006), 025
doi:10.1088/1126-6708/2006/06/025

\bibitem{Cham1} A.~H.~Chamseddine,
\emph{``Topological Gauge Theory of Gravity in Five-dimensions and All Odd Dimensions,''}
Phys. Lett. B \textbf{233}, 291-294 (1989).

\bibitem{Cham2} A.~H.~Chamseddine,
\emph{``Topological gravity and supergravity in various dimensions,''}
Nucl. Phys. B \textbf{346}, 213-234 (1990). 

\bibitem{Banados:2005rz}
M.~Banados, R.~Olea and S.~Theisen,
\emph{``Counterterms and dual holographic anomalies in CS gravity,''}
JHEP \textbf{10} (2005), 067
doi:10.1088/1126-6708/2005/10/067

\bibitem{Cvetkovic:2017fxa}
B.~Cvetkovi\'c, O.~Miskovic and D.~Simi\'c,
\emph{``Holography in Lovelock Chern-Simons AdS Gravity,''}
Phys. Rev. D \textbf{96} (2017) no.4, 044027
doi:10.1103/PhysRevD.96.044027

\bibitem{Gallegos:2020otk}
A.~D.~Gallegos and U.~G\"ursoy,
\emph{``Holographic spin liquids and Lovelock Chern-Simons gravity,''}
JHEP \textbf{11} (2020), 151
doi:10.1007/JHEP11(2020)151

\bibitem{Mignemi:1998us}
S.~Mignemi,
\emph{``The Dynamical structure of four-dimensional Chamseddine's gauge theory of gravity,''}
Phys. Rev. D \textbf{59} (1999), 064022
doi:10.1103/PhysRevD.59.064022

\bibitem{Ebert:2022ehb}
S.~Ebert, C.~Ferko, H.~Y.~Sun and Z.~Sun,
\emph{``$T\bar{T}$ in JT Gravity and BF Gauge Theory,''}
SciPost Phys. \textbf{13} (2022) no.4, 096
doi:10.21468/SciPostPhys.13.4.096

\bibitem{Dordevic:2022ruk}
D.~\DJ{}or\dj{}evi\'c and D.~Go\v{c}anin,
\emph{``Noncommutative $D=5$ Chern\textendash{}Simons gravity: Kaluza\textendash{}Klein reduction and chiral gravitational anomaly,''}
Eur. Phys. J. C \textbf{82} (2022) no.8, 672
doi:10.1140/epjc/s10052-022-10657-7


\bibitem{Aksteiner:2015uxw}
S.~Aksteiner and Y.~Korovin,
\emph{``New modes from higher curvature corrections in holography,''}
JHEP \textbf{03} (2016), 166
doi:10.1007/JHEP03(2016)166


\bibitem{Iliesiu:2019xuh} L.~V.~Iliesiu, S.~S.~Pufu, H.~Verlinde and Y.~Wang, \emph{``An exact quantization of Jackiw-Teitelboim gravity} JHEP \textbf{11} (2019), 091 doi:10.1007/JHEP11(2019)091

\bibitem{Erdmenger:2022nhz}
J.~Erdmenger, B.~He\ss{}, I.~Matthaiakakis and R.~Meyer,
\emph{``Universal Gibbons-Hawking-York term for theories with curvature, torsion and non-metricity,''} 
SciPost Phys. \textbf{14} (2023), 099
doi:10.21468/SciPostPhys.14.5.099


\bibitem{MacDowell-Mansouri}
S.~W.~MacDowell and F.~Mansouri,
\emph{``Unified Geometric Theory of Gravity and Supergravity,''}
Phys. Rev. Lett. \textbf{38}, 739 (1977). https://doi.org/10.1103/PhysRevLett.38.739

\bibitem{stelle-west} K.~S.~Stelle and P.~C.~West,
\emph{``Spontaneously Broken De Sitter Symmetry and the Gravitational Holonomy Group,''}
Phys. Rev. D \textbf{21}, 1466 (1980). https://doi.org/10.1103/PhysRevD.21.1466

\bibitem{Canfora:2007xs}
F.~Canfora, A.~Giacomini and R.~Troncoso,
\emph{``Black holes, parallelizable horizons and half-BPS states for the Einstein-Gauss-Bonnet theory in five dimensions,''}
Phys. Rev. D \textbf{77} (2008), 024002
doi:10.1103/PhysRevD.77.024002

\bibitem{Brihaye:2013vsa}
Y.~Brihaye and E.~Radu,
\emph{``Black hole solutions in d=5 Chern-Simons gravity,''}
JHEP \textbf{11} (2013), 049
doi:10.1007/JHEP11(2013)049

\bibitem{Tetradis:2011jn}
N.~Tetradis,
\emph{``Entropy from AdS(3)/CFT(2),''}
JHEP \textbf{02} (2012), 054
doi:10.1007/JHEP02(2012)054

\bibitem{Gursoy:2007cb}
U.~Gursoy and E.~Kiritsis,
\emph{``Exploring improved holographic theories for QCD: Part I,''}
JHEP \textbf{02} (2008), 032
doi:10.1088/1126-6708/2008/02/032

\bibitem{Gursoy:2009jd}
U.~Gursoy, E.~Kiritsis, L.~Mazzanti and F.~Nitti,
\emph{``Improved Holographic Yang-Mills at Finite Temperature: Comparison with Data,''}
Nucl. Phys. B \textbf{820} (2009), 148-177
doi:10.1016/j.nuclphysb.2009.05.017

\bibitem{Oliva:2010xn}
J.~Oliva and R.~Troncoso,
\emph{``Exact quasinormal modes for a special class of black holes,''}
Phys. Rev. D \textbf{82} (2010), 027502
doi:10.1103/PhysRevD.82.027502

\bibitem{Kanitscheider:2009as}
I.~Kanitscheider and K.~Skenderis,
\emph{``Universal hydrodynamics of non-conformal branes,''}
JHEP \textbf{04} (2009), 062
doi:10.1088/1126-6708/2009/04/062


\bibitem{Henningson:1998gx}
M.~Henningson and K.~Skenderis,
\emph{``The Holographic Weyl anomaly,''}
JHEP \textbf{07} (1998), 023
doi:10.1088/1126-6708/1998/07/023


\bibitem{Taylor:2017dly}
M.~Taylor,
\emph{``Generalized conformal structure, dilaton gravity and SYK,''}
JHEP \textbf{01} (2018), 010
doi:10.1007/JHEP01(2018)010

\bibitem{Banados:2004zt}
M.~Banados, A.~Schwimmer and S.~Theisen,
\emph{``Chern-Simons gravity and holographic anomalies,''}
JHEP \textbf{05}, 039 (2004)
doi:10.1088/1126-6708/2004/05/039

\bibitem{Isler:1989hq}
K.~Isler and C.~A.~Trugenberger,
\emph{``A Gauge Theory of Two-dimensional Quantum Gravity,''}
Phys. Rev. Lett. \textbf{63} (1989), 834
doi:10.1103/PhysRevLett.63.834

\bibitem{Chamseddine:1989wn}
A.~H.~Chamseddine and D.~Wyler,
\emph{``Topological Gravity in (1+1)-dimensions,''}
Nucl. Phys. B \textbf{340} (1990), 595-616
doi:10.1016/0550-3213(90)90460-U

\bibitem{Almheiri:2014cka}
A.~Almheiri and J.~Polchinski,
\emph{``Models of AdS$_{2}$ backreaction and holography,''}
JHEP \textbf{11} (2015), 014
doi:10.1007/JHEP11(2015)014


\bibitem{Witten:2020ert}
E.~Witten,
\emph{``Deformations of JT Gravity and Phase Transitions,''}
[arXiv:2006.03494 [hep-th]].

\bibitem{Lemos:1996bq}
J.~P.~S.~Lemos,
\emph{``Thermodynamics of the two-dimensional black hole in the Teitelboim-Jackiw theory,''}
Phys. Rev. D \textbf{54} (1996), 6206-6212
doi:10.1103/PhysRevD.54.6206


\bibitem{DHoker:2019clx}
E.~D'Hoker and P.~Kraus,
\emph{``Gravitational Wilson lines in AdS$_3$,''}
Roman Jackiw, WORLD SCIENTIFIC
doi:10.1142/9789811210679\_0014

\bibitem{Bhatta:2016hpz}
A.~Bhatta, P.~Raman and N.~V.~Suryanarayana,
\emph{``Holographic Conformal Partial Waves as Gravitational Open Wilson Networks,''}
JHEP \textbf{06} (2016), 119
doi:10.1007/JHEP06(2016)119

\bibitem{Ammon:2013hba}
M.~Ammon, A.~Castro and N.~Iqbal,
\emph{``Wilson Lines and Entanglement Entropy in Higher Spin Gravity,''}
JHEP \textbf{10} (2013), 110
doi:10.1007/JHEP10(2013)110

\bibitem{Freidel:2006hv}
L.~Freidel, J.~Kowalski-Glikman and A.~Starodubtsev,
\emph{``Particles as Wilson lines of gravitational field,''
Phys. Rev. D \textbf{74} (2006), 084002}
doi:10.1103/PhysRevD.74.084002

\bibitem{Fairbairn:2008hy}
W.~J.~Fairbairn,
\emph{``On gravitational defects, particles and strings,''}
JHEP \textbf{09} (2008), 126
doi:10.1088/1126-6708/2008/09/126

\bibitem{Erdmenger:2017pfh}
J.~Erdmenger and N.~Miekley,
\emph{``Non-local observables at finite temperature in AdS/CFT,''}
JHEP \textbf{03} (2018), 034
doi:10.1007/JHEP03(2018)034

\bibitem{geod}
G.~Gianluca, and G.~Zampieri. 
\emph{''The Geodesics for Poincaré’s Half-Plane: A Nonstandard Derivation''} 
The American Mathematical Monthly 130.5 (2023): 478-481.
doi.org/10.1080/00029890.2023.2181574



\bibitem{Almheiri:2016blp}
A.~Almheiri, X.~Dong and B.~Swingle,
\emph{``Linearity of Holographic Entanglement Entropy,''}
JHEP \textbf{02} (2017), 074
doi:10.1007/JHEP02(2017)074

\bibitem{Papadodimas:2015jra}
K.~Papadodimas and S.~Raju,
\emph{``Remarks on the necessity and implications of state-dependence in the black hole interior,''}
Phys. Rev. D \textbf{93} (2016) no.8, 084049
doi:10.1103/PhysRevD.93.084049

\bibitem{Erdmenger:2023hne}
J.~Erdmenger, B.~He\ss{}, I.~Matthaiakakis and R.~Meyer,
\emph{``Gibbons-Hawking-York boundary terms and the generalized geometrical trinity of gravity,''}
[arXiv:2304.06752 [hep-th]].

\bibitem{Inostroza:2014vua}
C.~Inostroza, A.~Salazar and P.~Salgado,
\emph{``Brans\textendash{}Dicke gravity theory from topological gravity,''}
Phys. Lett. B \textbf{734} (2014), 377-382
doi:10.1016/j.physletb.2014.05.080

\bibitem{Loran:2013fca}
F.~Loran,
\emph{``Holographic anomaly in 3d $f$(Ric) gravity,''}
Class. Quant. Grav. \textbf{30} (2013), 195010
doi:10.1088/0264-9381/30/19/195010

\bibitem{Hashimoto:2013bna}
K.~Hashimoto, N.~Iizuka and T.~Kimura,
\emph{``Towards Holographic Spintronics,''}
Phys. Rev. D \textbf{91} (2015) no.8, 086003
doi:10.1103/PhysRevD.91.086003




\bibitem{Blagojevic:2013bu}
M.~Blagojevic, B.~Cvetkovic, O.~Miskovic and R.~Olea,
\emph{``Holography in 3D AdS gravity with torsion,''}
JHEP \textbf{05} (2013), 103
doi:10.1007/JHEP05(2013)103



\end{thebibliography}
\end{document}